\documentclass[aps,prb,preprint,groupedaddress]{revtex4}

\usepackage[draft]{graphicx}
\usepackage{bm}

\begin{document}


\title{Inelastic neutron scattering study of phonons and magnetic excitations in LaCoO$_{3}$}


\author{Y. Kobayashi}
 \email{koba@pc.uec.ac.jp}
 \affiliation{Department of Applied Physics and Chemistry, The University of Electro-Communications, Chofu, Tokyo 182-8585, Japan}
\author{Thant Sin Naing}
\affiliation{Department of Applied Physics and Chemistry, The University of Electro-Communications, Chofu, Tokyo 182-8585, Japan}
\author{M. Suzuki}
\affiliation{Department of Applied Physics and Chemistry, The University of Electro-Communications, Chofu, Tokyo 182-8585, Japan}
\author{M. Akimitsu}
\affiliation{Department of Applied Physics and Chemistry, The University of Electro-Communications, Chofu, Tokyo 182-8585, Japan}
\author{K. Asai}
\affiliation{Department of Applied Physics and Chemistry, The University of Electro-Communications, Chofu, Tokyo 182-8585, Japan}
\author{K. Yamada}
\affiliation{Institute for Materials Research, Tohoku University, Sendai, 980-8577, Japan}
\author{J. Akimitsu}
\affiliation{Department of Physics and Mathematics, Aoyama Gakuin University, Sagamihara 229-8551, Japan}
\author{P. Manuel}
\affiliation{ISIS Facility, CCLRC Rutherford Appleton Laboratory, Didcot OX110QX, UK}
\author{J. M. Tranquada and G. Shirane}
\affiliation{Brookhaven National Laboratory, New York 11973 USA}
%

\date{\today}
\begin{abstract}
We have investigated the phonon and the magnetic excitations in 
LaCoO$_{3}$ by inelastic neutron scattering measurements.  The acoustic
phonon dispersions show some characteristic features of the folded
Brillouin zone (BZ) for the rhombohedrally distorted perovskite structure
containing two chemical formula units of LaCoO$_{3}$ in the unit cell. 
We observed two transverse optical (TO) phonon branches along ($\delta$
$\delta$ $\delta$), consistent with previously reported Raman active
$E_{g}$ modes which show remarkable softening associated with the
spin-state transition [Ishikawa \textit{et al.}, (Phys. Rev. Lett. 93
(2004) 136401.)].  We found that the softening takes place in the TO mode
over the whole BZ.  In contrast, the acoustic phonons show no anomalous
softening associated with the spin-state transition.
The low-energy paramagnetic scattering at 8~K is weak, increasing towards
a maximum at $E \gtrsim 15$ meV, consistent with
excitation of the nonmagnetic low-spin to magnetic intermediate-spin
state of Co$^{3+}$ ions.
\end{abstract}


\maketitle

\section{Introduction}

Lanthanum cobalt oxide LaCoO$_{3}$ exhibits magnetic-electronic 
transitions around 100 and 500 K.
\cite{Heikes,Raccah,Asai1,Asai2,Asai4,Asai3}  In order to explain these
two transitions consistently, a two-stage spin-state transition model has
been proposed.  Within the model, the 100 K transition is associated with
the thermal excitation of Co$^{3+}$ ions from the low spin (LS; $S=0$)
ground state to an intermediate spin (IS; $S=1$) state, whereas the 500 K
transition corresponds to a crossover from the IS state to a mixed state
of IS and the high spin (HS; $S=2$) state.\cite{Asai3}  
Reducing the volume either by hydrostatic pressure\cite{Asai4} or by
subsitution of rare-earth elements,\cite{baie05,nekr03} with a smaller
ionic radius, for La causes the spin-state transitions to move to higher
temperature, and makes the two transitions less distiguishable.  Korotin 
\textit{et al.}\cite{Korotin} were the first to show theoretically that
the IS state is energetically close to the LS state due to a large
hybridization between Co-$3d$ and O-$2p$ orbitals. 
In a purely ionic picture, however,
the ground state of Co$^{3+}$ ions is either the LS state ($\Delta_{cf} >
\Delta_{ex}$) or the HS state ($\Delta_{cf} < \Delta_{ex}$), where
$\Delta_{cf}$ and $\Delta_{ex}$ are the crystal-field splitting and the
intra-atomic exchange interaction, respectively.
\cite{Tanabe,Raccah,Korotin}  
Recently, it has been argued that the IS state might be
understood by including spin-orbit effects in the purely ionic
model,\cite{radw99} and comparisons have been made with
electron-spin-resonance measurements.\cite{Noguchi}
Thus, the nature of the IS-state Co$^{3+}$ is still controversial.  

The 100-K spin-state transition couples with phonons.  In fact, lattice 
anomalies correlated with the spin-state transition have been reported. 
The lattice volume shows an anomalous expansion of about 1\%\ around 100
K. \cite{Asai3}  The elastic constant measured by longitudinal
ultrasound shows a dip of about 30\%\ in magnitude around 100 K,
suggesting a softening associated with coexistence of the LS and the
IS-states of Co$^{3+}$. \cite{Murata}  In addition, it is expected that
the hybridization-induced IS-state of the Co$^{3+}$ ion should have an
associated Jahn-Teller (JT) distortion, since the $e_{g}$ orbital is
partially occupied.  Several experiments suggest the orbital ordering due
to the JT distortion. Yamaguchi \textit{et al.}\cite{Yamaguchi} have
reported an anomalous temperature dependence of the intensity and the
splitting of the phonon modes with the energies of 35, 50 and 73 meV
based on an infrared spectroscopic study, and proposed a local JT
distortion resulting from the appearance of the IS-state Co$^{3+}$. 
Louca and Sarrao have proposed a local static JT distortion based on the
pair-density-function analysis of the neutron scattering in the
paramagnetic insulating phase of La$_{1-x}$Sr$_{x}$CoO$_{3}$.
\cite{Louca}  A structural analysis using synchrotron X-ray diffraction,
performed by Maris \textit{et al.},
\cite{Maris} infers an alternation of short and long Co-O bonds
accompanied with a subsequent change from rhombohedral $R\bar{3}c$ to
monoclinic $I2/a$ for 20--300 K, suggesting the presence of the
$e_{g}$ orbital ordering on the IS-state Co$^{3+}$ with a cooperative JT
distortion of the $Q_{2}$ type. \cite{Kanamori}  The JT distortion
increases sharply around 70 K with increasing temperature.  Recently,
Ishikawa \textit{et al.} have reported an energy shift and broadening
associated with the population of the IS-state Co$^{3+}$ ions for the
phonons of the $E_{g}$ rotational mode of O atoms at 11 meV, the $E_{g}$
vibration mode of La atoms at 21 meV, and the $A_{1g}$ rotational mode of
O atoms at 32 meV (energies are the values at 5 K) based on the Raman
scattering experiment. \cite{Ishikawa}  In addition, they have reported
the appearance of the satellite peaks associated with the IS-state at the
energies of 19, 45 and 84 meV with increasing temperature.  Based on the
results, they supported the orbital ordering due to the cooperative JT
distortion proposed by Maris \textit{et al.} \cite{Maris}  These studies
show that the investigation of the phonons is of paramount importance in
understanding the spin-state transition.

It is important to clarify the phonon dispersion over the Brillouin zone 
in order to investigate the coupling between the phonons and the
spin-state transition.  The previous studies suggesting the lattice
anomaly \cite{Asai3, Murata, Yamaguchi, Louca, Maris, Ishikawa} probe
only the zone center ($\Gamma$-point) of the Brillouin
zone.  Only neutron scattering experiments using a single crystalline
sample affords information over the Brillouin zone.  Louca \textit{et
al.} have reported the $\bm{q}$-integrated phonon spectra by neutron
scattering study but they showed no information on the $\bm{q}$
dependence of the phonons. \cite{Louca}  The phonon dispersion in the
rhombohedral phase of the perovskite oxide over the whole Brillouin zone
has not been reported to our knowledge although many studies of the
soft-phonon mode near the structural phase transition in the cubic phase
of LaAlO$_{3}$ have been reported. \cite{Axe, Kiems} 

In this paper, we report the inelastic neutron scattering study on 
LaCoO$_{3}$ single crystal in order to determine the phonon dispersion
and find the phonon anomaly associated with the spin-state transition of
LaCoO$_{3}$.  In addition to the phonon scattering, we have investigated
the inelastic magnetic neutron scattering associated with the energy gap
between the LS and the IS state, which has been reported to be about 16
meV.\cite{Kobayashi1,Noguchi}

\section{Sample preparation and experimental procedure}

Single crystals of LaCoO$_{3}$ were grown with a lamp-image floating zone 
furnace by melting polycrystalline samples prepared by a
solid-state-reaction of predried La$_{2}$O$_{3}$ and CoO.
\cite{Kobayashi1}  Laue photographs confirmed that each sample is a proper
single crystal.  LaCoO$_{3}$ has a rhombohedrally
distorted perovskite structure with the crystal space group of
$R\bar{3}c$ which contains two chemical formula units.  Since the crystal
contains four twins, each of which has the principal axis parallel to one
of the $\langle 111 \rangle$-axes of the pseudo-cubic unit cell, and the
rhombohedral distortion of the crystal from the cubic structure is small,
we use the crystallographic indices based on the pseudo-cubic cell ($a =
3.83$ \AA \ and $\alpha = 90.6^\circ$ at 295 K \cite{Thornton2})
containing one molecule of LaCoO$_{3}$.  We must mention that the R-point
of the pseudo-cubic Brillouin zone (0.5 0.5 0.5) is the zone center
($\Gamma$-point) of the rhombohedral one due to the Brillouin zone
folding.  A rod of the single crystal with about 5 mm in diameter and
about 40 mm in length and an assembly of similar four rods were used for
the measurement at Brookhaven National Laboratory (BNL) and JAERI,
respectively.  The crystals were oriented with their $[01\bar{1}]$
directions vertical.  The misalignment of the four crystals with respect
to each other is within 1$^\circ$.  

The inelastic neutron scattering measurements on a triple-axis 
spectrometer (TAS) H7 at BNL were performed in 1996.  After preliminary
measurements on PRISMA at ISIS in Rutherford Appleton Laboratory in 2003,
the detailed inelastic neutron scattering experiments were performed on a
triple-axis spectrometer TOPAN at JRR-3M in JAERI, Tokai.  Both TAS
experiments were performed with fixed final neutron energies ($E_f$) of
30.5 and 14.7 meV, but the horizontal collimation was 40'-40'-40'-80' for
the BNL measurement and 30'-100'-60'-blank and blank-60'-60'-60',
respectively for the JAERI measurement.  For the phonon scattering, the
scans were performed to measure both the transverse and the longitudinal
phonons with the propagating vectors $(\delta, \delta, \delta)$,
$(\delta, 0, 0)$ and $(0, \delta, \delta)$.  For the magnetic excitation
study, we have performed inelastic neutron scattering measurements around
(1 0 0) and around an equivalent reciprocal lattice point (3 0 0). 

\section{Experimental Results}

\subsection{Phonon scattering}

\begin{figure}
\includegraphics[width=8cm,height=5cm,clip]{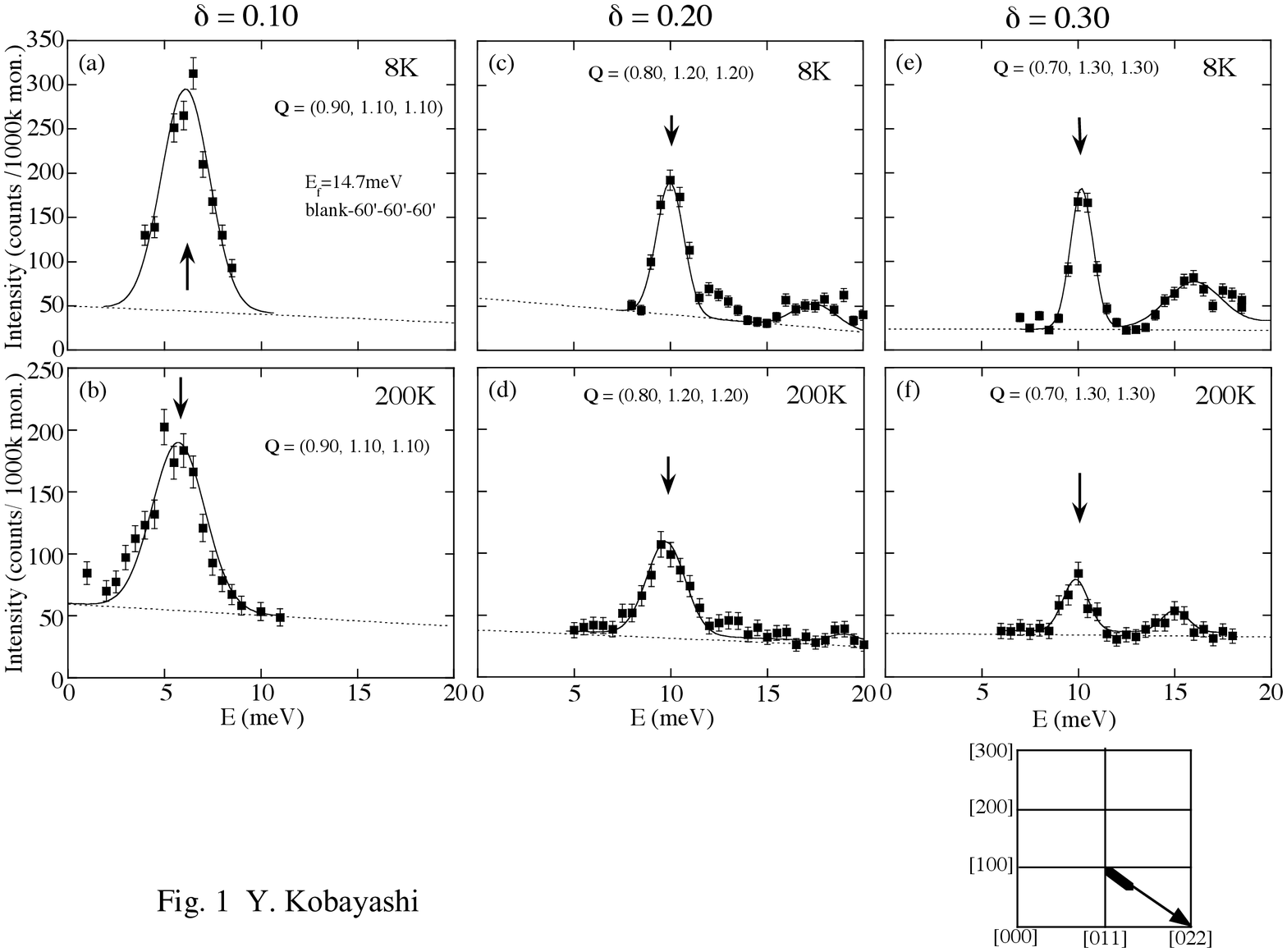}%
\caption{\label{fig:1}  The inelastic neutron scattering intensities per 
1000 k monitor for LaCoO$_{3}$ at $\bm{Q} = (1-\delta, 1+\delta,
1+\delta)$ at $T = 8$ and 200 K with $E_f = 14.7$ meV, after being
corrected by the temperature factor.  The solid lines are fits to the
data using a Gaussian line shape.  The arrows indicate the peak position
of the acoustic phonons.  The dotted lines are the background level.}
\end{figure}

The inelastic neutron scattering intensities at $\bm{Q} = (1-\delta,
1+\delta, 1+\delta)$ are shown in Fig.~\ref{fig:1}.  The data were
taken with the fixed final neutron energy ($E_f$) of 14.7 meV at
temperatures ($T$) of 8 and 200 K.  The intensities have been corrected
by the temperature factor $\langle n + 1\rangle \equiv 1/[1 - \exp(- E /
k_{B}T)]$.  Relatively sharp peaks are observed at about 6 meV for
$\delta$ = 0.1 and about 10 meV for $\delta$ = 0.2 and 0.3.  We identify
these phonons as the transverse acoustic (TA) phonons propagating along
$\bm{q}$ = $(\delta$, $\delta$, $\delta)$.  We confirmed these phonons
also at equivalent reciprocal lattice points $\bm{Q}$ = $(1+\delta$,
$1-\delta$, $1-\delta)$.  We find that the energy shift for the acoustic
phonons between 8 and 200 K is small for all $\delta$ although the energy
width at 200 K is broader than that at 8 K.  The broad peaks observed in
the higher energy region ($E$ $\sim$ 15 meV) for $\delta$ = 0.2 and 0.3
are due to the optical phonon scattering described later.  We cannot
identify the small peak observed at around 12 meV for $\delta$ = 0.2.  We
performed a similar measurement along $\bm{Q}$ = $(1+\delta$, $1+\delta$,
$1+\delta)$ and observed the longitudinal acoustic (LA) phonons
propagating along $\bm{q}$ = $(\delta$, $\delta$, $\delta)$ (not shown).

\begin{figure}
\includegraphics[width=8cm,height=5cm,clip]{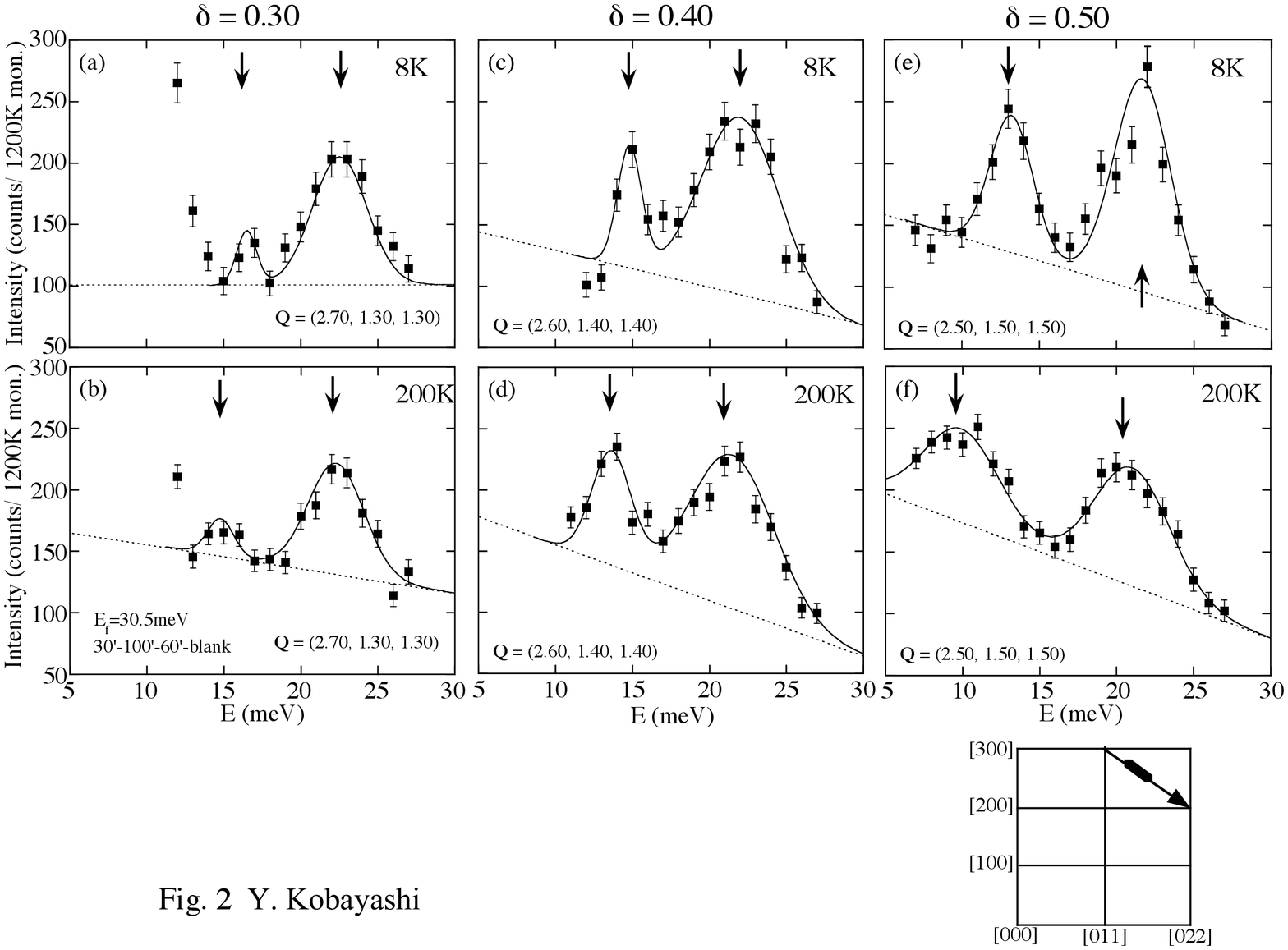}%
\caption{\label{fig:2}   The inelastic neutron scattering intensities per 
1200 k monitor at $\bm{Q} = (3-\delta, 1+\delta, 1+\delta)$ at $T
= 8$ and 200 K with $E_f = 30.5$ meV, after being corrected by the
temperature factor.  The solid lines are fits to the data using a
Gaussian line shape.  The arrows indicate the peak position of the
optical phonons.  The dotted lines are the background level.}
\end{figure}

The inelastic neutron scattering intensities for the transverse optical 
phonons at $T$ = 8 and 200 K are shown in Fig.~\ref{fig:2}.    The data
were taken at $\bm{Q} = (3-\delta, 1+\delta, 1+\delta)$ with $E_f = 30.5$
meV.  At 8K at $\bm{Q} = (2.5, 1.5, 1.5)$ ($\delta = 0.5$), two broad
peaks were observed at around $E = 13$ and 22 meV.   We confirmed that
the two peaks are due to the phonon scattering by comparing  the
scattering intensities measured at other equivalent reciprocal lattice
points  [$\bm{Q} = (1.5, 0.5, 0.5)$ and (2.5, 1.5, 1.5)].   We consider
that these two phonons at the R-points for the pseudo-cubic cell are
identical to the phonons of the $E_{g}$ rotational mode of O atoms (11
meV at 5 K) and $E_{g}$ vibration mode of La atoms (21 meV at 5 K)
observed in a Raman scattering experiment by Ishikawa \textit{et al.}
\cite{Ishikawa} although the latter phonon is slightly different in
energy. It should be noted that the Raman active modes for the
rhombohedral $R\bar{3}c$ symmetry are at the R-point of the pseudo-cubic
Brillouin zone since the atomic displacements of the modes are anti-phase
for the nearest neighbor pairs along the [111]-axis.
\cite{Ishikawa,Granado,Abrashev}

\begin{figure}
\includegraphics[width=8cm,height=5cm,clip]{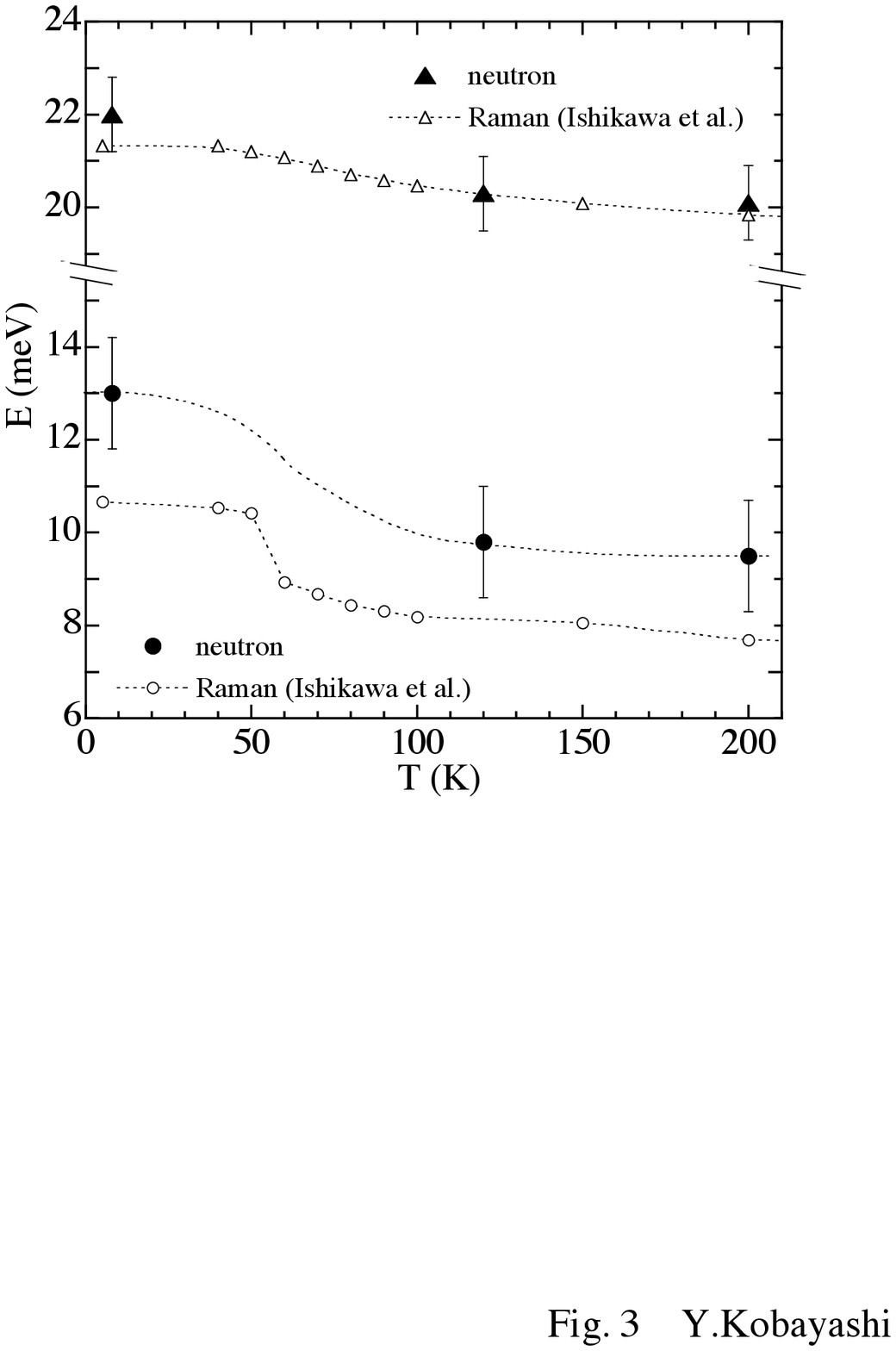}%
\caption{\label{fig:3}  Temperature dependence of the phonon energy at 
$\bm{Q} = (2.50, 1.50, 1.50)$.  The Raman data are from Ref.\
\onlinecite{Ishikawa}.  The dashed lines are guides to eyes.}  
\end{figure}

The optical phonons at $\delta = 0.5$ show a remarkable softening and
broadening with increasing temperature from 8 K to 200 K. We measured the
phonon spectra also at 120 K and found that the optical phonon spectra
are almost identical to those at 200 K.  The temperature dependence of
the phonon energy is in agreement with the result of the Raman
spectroscopy (see Fig.~\ref{fig:3}).   The broadening of the phonon peak
of the O-rotational mode with increasing temperature is also consistent
with the result of Raman scattering \cite{Ishikawa}.

\begin{figure}
\includegraphics[width=8cm,height=5cm,clip]{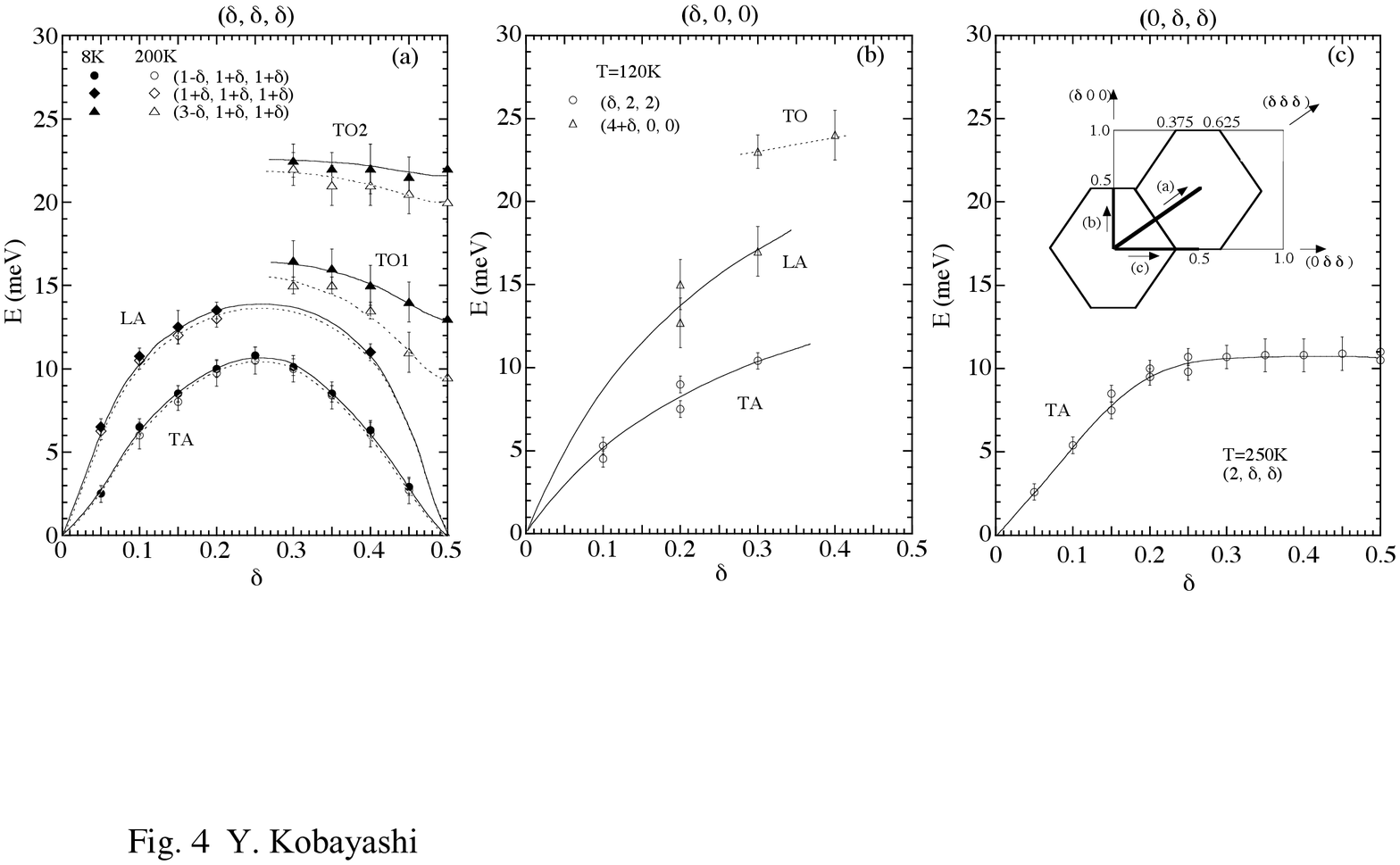}%
\caption{\label{fig:4}  The phonon dispersion curves of LaCoO$_{3}$ along 
(a) ($\delta \ \delta \ \delta$), (b) (0 $\delta$ $\delta$) and (c)
($\delta$ 0 0).  The lines are guides to eyes.  Inset in Fig.~\ref{fig:4}
(c) shows the trajectories for the scans.  The hexagon and the rectangle
are the $(01\bar{1})_{c}$ section of the Brillouin zones for the
rhombohedral and the pseudo-cubic unit cells, respectively.}
\end{figure}

The phonon dispersion along $\bm{q} = (\delta, \delta, \delta)$
obtained in the present experiment is summarized in Fig.~\ref{fig:4}
(a).  In addition to the LA and the TA branches, we see two transverse
optical branches; the lower one continuing to the $E_{g}$ rotational mode
of O atoms and the other to the $E_{g}$ vibration mode of La atoms at
$\delta = 0.5$. \cite{Ishikawa,Granado,Abrashev}  We name the former and
the latter branches TO1 and TO2, respectively.  The phonon energies of
both TO1 and TO2 remarkably decrease with increasing temperature from 8 K
to 200 K.  The softening is most pronounced at $\delta$ = 0.5, but it
is not limited to that point.  In contrast to the optical
phonons, the softening in both LA and TA branches is negligible in all
of the reciprocal lattice space investigated.  

We measured the phonon scattering at $\bm{Q}$ = ($\delta$ 2 2) and 
(4+$\delta$ 0 0) with $E_{f}$ = 30.5 meV at 120 K to investigate the
phonon dispersion along ($\delta$ 0 0) for the TA and the LA modes, and
$\bm{Q}$ = (2 $\delta$ $\delta$) with $E_{f}$ = 14.7 meV at 250 K for the
TA phonon mode along (0 $\delta \ \delta$).   The phonon dispersion curves
are shown in Figs.~\ref{fig:4} (b) and (c), respectively.  The
trajectories of the $\bm{q}$ in the reduced Brillouin zone for the
rhombohedral crystal structure ($R\bar{3}c$) containing two chemical
formula units of LaCoO$_{3}$ are shown in the inset of Fig.~\ref{fig:4}
(c).

\subsection{Magnetic scattering}

\begin{figure}
\includegraphics[width=8cm,height=5cm,clip]{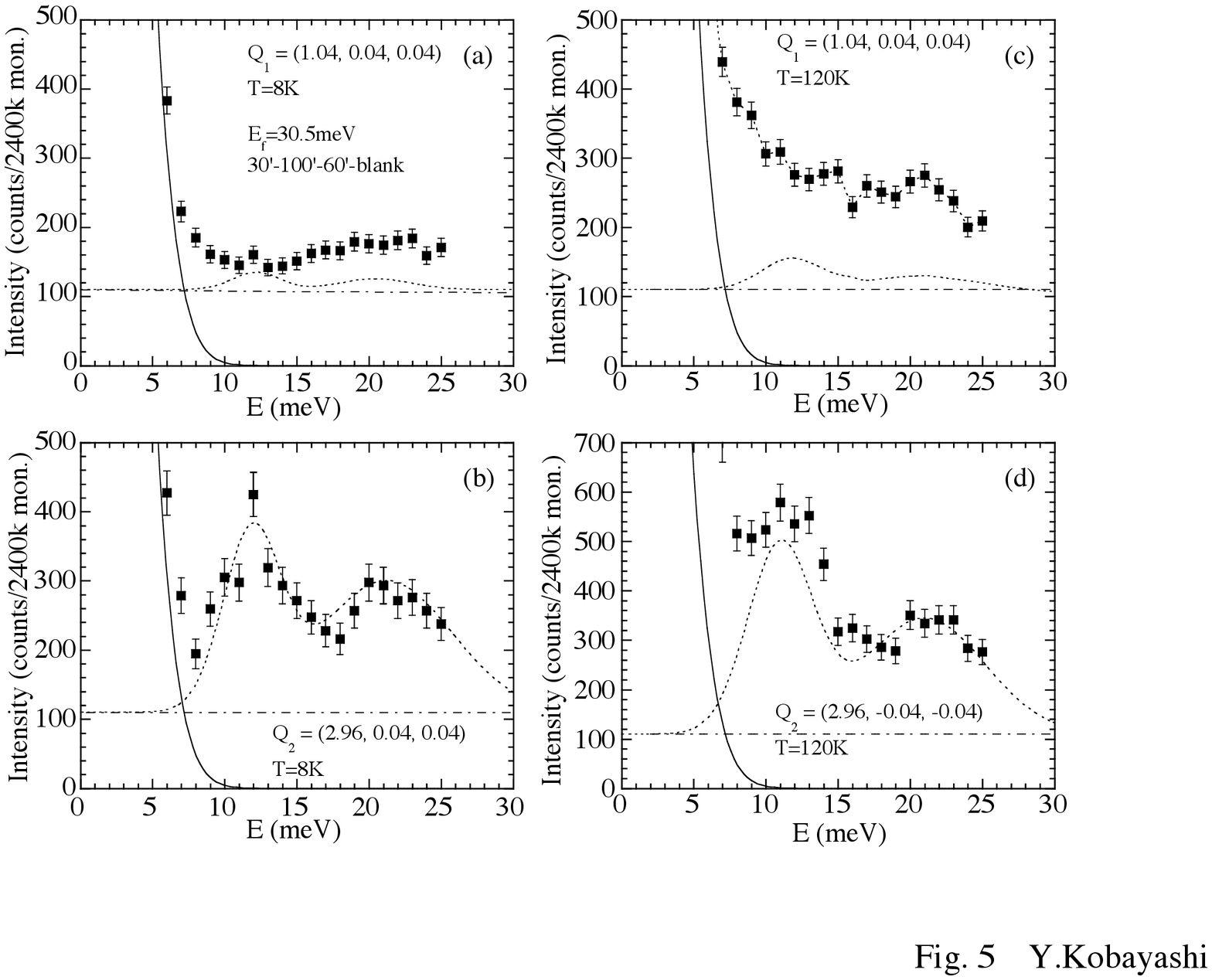}%
\caption{\label{fig:5}  The energy spectra of the inelastic neutron 
scattering intensities at $\bm{Q}_{1}$ = (1.04, 0.04, 0.04) (a) at 8 and
(b) 120 K, and $\bm{Q}_{2}$ = (2.96, 0.04, 0.04) (c) at 8 and (d) 120 K. 
The solid lines are the Bragg (or some incoherent) scattering tail.  The
dotted lines are the phonon component including the background intensity
shown by the dash-dotted lines.}  
\end{figure}

The energy-dependent spectra at $\bm{Q}_{1}$ = (1.04, 0.04, 0.04) and an
equivalent  reciprocal lattice point $\bm{Q}_{2}$ = (2.96, 0.04, 0.04)
measured at temperatures 8 and 120 K are shown in Fig.~\ref{fig:5}.
We expect the signal at $\bm{Q}_{1}$ to be dominated by magnetic
scattering and that at $\bm{Q}_{2}$ to be dominated by phonons.  Let us
consider first the $\bm{Q}_{2}$ data, in Fig.~\ref{fig:5}(b) and (d).  To
fit these spectra, we assume that the scattering can be decomposed into
three components: 1) gaussian tail from elastic scattering, 2) the two
gaussian phonon peaks, and 3) a temperature- and energy-independent
background.  The magnetic scattering at $\bm{Q}_{2}$ should be small
due to the magnetic form factor.   In fitting the data, the background is
constrained by the assumption that the temperature dependence of the
phonon scattering is due to entirely to the Bose factor.  The various
fitted components are indicated in the figure.

\begin{figure}
\includegraphics[width=8cm,height=5cm,clip]{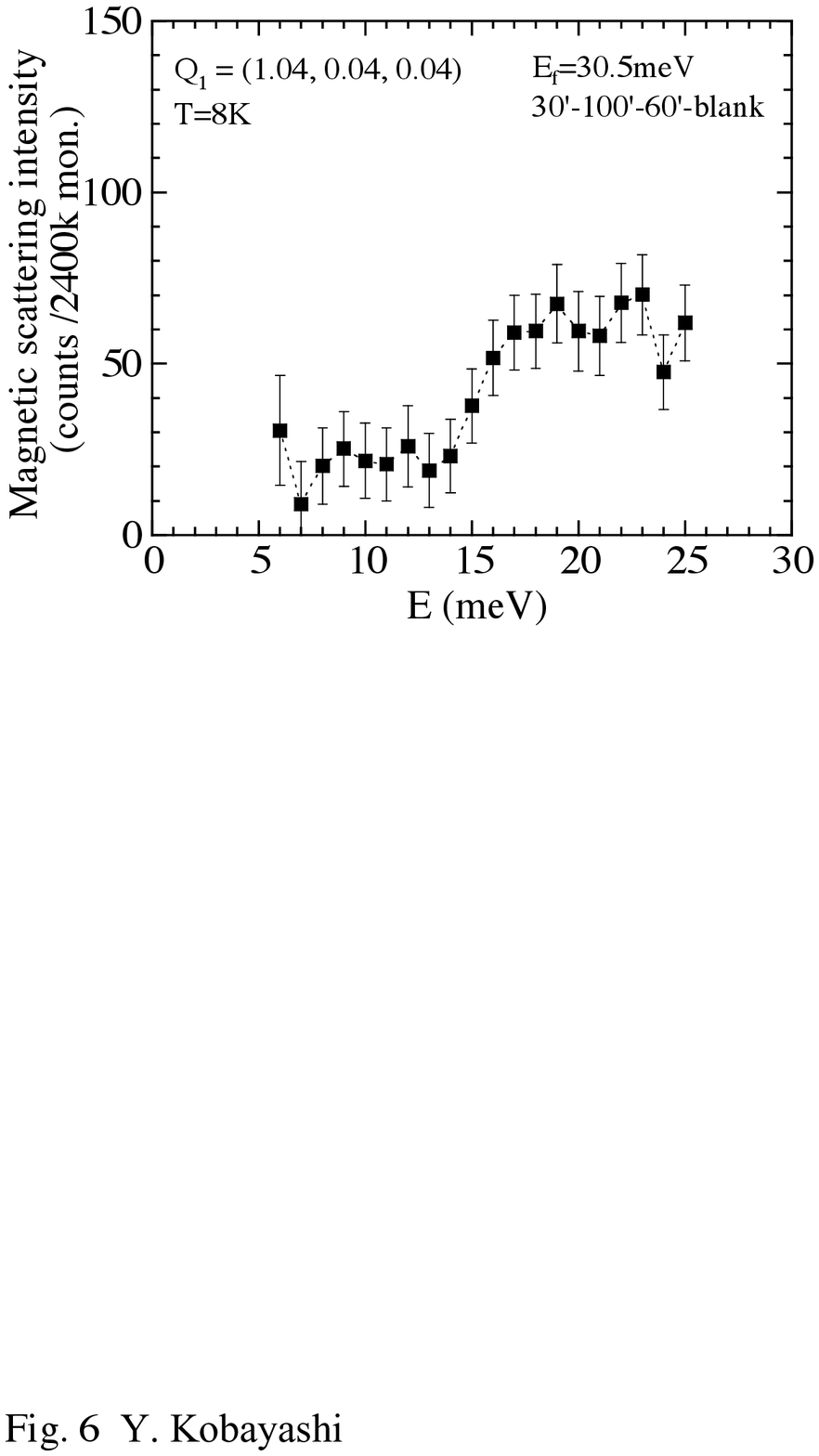}%
\caption{\label{fig:6}  The magnetic scattering intensity at 
$\bm{Q}_{1}$ = (1.04, 0.04, 0.04) at 8 K.}  
\end{figure}

To extract the magnetic scattering from the spectra taken at
$\bm{Q}_{1}$, we assume that the elastic tail and the background are the
same as those determined at $\bm{Q}_{2}$.  The phonon intensity is
estimated by assuming that it varies as $|\bm{Q}|^2$.  The components so
obtained are indicated in Fig.~\ref{fig:5}(a) and (c).  We assume that
the difference between the measured data and the sum of the three
components (tail, background, and phonons) is magnetic scattering.  The
deduced magnetic scattering at 8~K is shown in Fig.~\ref{fig:6}.  The
significance of this signal, and the increased scattering at 120 K, will
be discussed in the next section.

\section{Discussion}

\subsection{Phonon scattering}

The features of the acoustic phonon dispersion along the principal axes of 
the reciprocal lattice space for the pseudo cubic unit cell are as
follows.  (1) For $\bm{q}$ = ($\delta$ $\delta$ $\delta$), both LA and TA
phonon branches are folded about $\delta$ = 0.25.  (2) For $\bm{q}$ = (0
$\delta$ $\delta$), the energy of the TA branch increases with increasing
$\delta$ for $\delta <$ 0.3, but levels off with $E \sim$ 10 meV for
larger $\delta$.  (3) For $\bm{q}$ = ($\delta$ 0 0), the energies of the
LA and TA branches increase monotonically with increasing $\delta$. 
The features (1) and (2) can be understood in terms of the reduced
Brillouin zone for the rhombohedral crystal structure of LaCoO$_{3}$ (see
the inset of Fig.~\ref{fig:4} (c)).  For $\bm{q}$ = ($\delta$ $\delta$
$\delta$), $\bm{q}$ crosses the zone boundary vertically at $\delta$ =
0.25, and the dispersion should be symmetric about this point.  
For $\bm{q}$ = (0 $\delta$ $\delta$), $\bm{q}$ arrives at the zone
boundary with $\delta$ = 0.375, and travels on the zone surface up to
$\delta$ = 0.5.  The result shows that the phonon energy on the
trajectory within the surface is not significant.  The dispersions along
($\delta$ $\delta$ $\delta$) and (0 $\delta$ $\delta$) are very different
from those in the cubic phase of SrTiO$_{3}$. \cite{Stirling}

It has been reported that the longitudinal sound velocity along [111] is 
more than 10 \% smaller at 200 K than at 8 K. \cite{Murata}  However, the
temperature difference between 8 and 200 K of the phonon energy found in
the present study is substantially smaller than that (at most 5\%) for
both the LA and the TA modes along ($\delta$ $\delta$ $\delta$) in the
whole Brillouin zone, which means that the softening found in the
ultrasonic measurements must be limited to a narrow region close to
$\delta$ = 0.

The optical phonons show remarkable softening with increasing temperature 
in accordance with the Raman spectroscopy results.\cite{Ishikawa}  A new
finding of the present study is that the softening occurs not only at
(0.5, 0.5, 0.5) but in the whole Brillouin zone along ($\delta$ $\delta$
$\delta)$ although it is most pronounced at $\delta$ = 0.5.  It should be
noted that the $E_{g}$ rotational mode of O atoms (TO1 at $\delta$ = 0.5)
and the $E_{g}$ vibration mode of La atoms (TO2 at $\delta$ = 0.5) show
substantial effects even though they are not especially sensitive to the
force constant between Co and O atoms.  More direct information about
possible JT distortions is contained in phonons such as the quadrupolar
mode found at much higher energies than those studied here.  It would be
quite interesting to probe that regime in a future study.

\subsection{Magnetic scattering}

The magnetic scattering at $\bm{Q}_{1}$ = (1.04, 0.04, 0.04) shown in 
Fig.~\ref{fig:6} 
is weak at low energy and suggests a broad peak at $\sim20$~meV. 
This energy scale is comparable to the energy
gap between the LS and the IS states deduced from
magnetization\cite{Asai4} (18 meV), lattice volume expansion\cite{Asai3}
(21 meV), NMR\cite{Kobayashi1} (15.5 meV), and ESR\cite{Noguchi} (12 meV)
measurements.  The similarity of these energies suggests that the
increment of the magnetic scattering is associated with the excitation
from the LS to the IS state of Co$^{3+}$ ions.  In other words, the
energy gap between the LS and the IS states are directly observed by the
present inelastic neutron scattering.  However, we regard this result as
tentative because the energy region about $\sim15$ meV is inevitably
contaminated by the optical phonons, and furthermore we cannot obtain
reliable temperature dependence of the magnitude of the excitation due to
the increasing paramagnetic scattering arising from the fluctuation of
the IS magnetic moments with increasing temperature.

\begin{figure}
\includegraphics[width=8cm,height=5cm,clip]{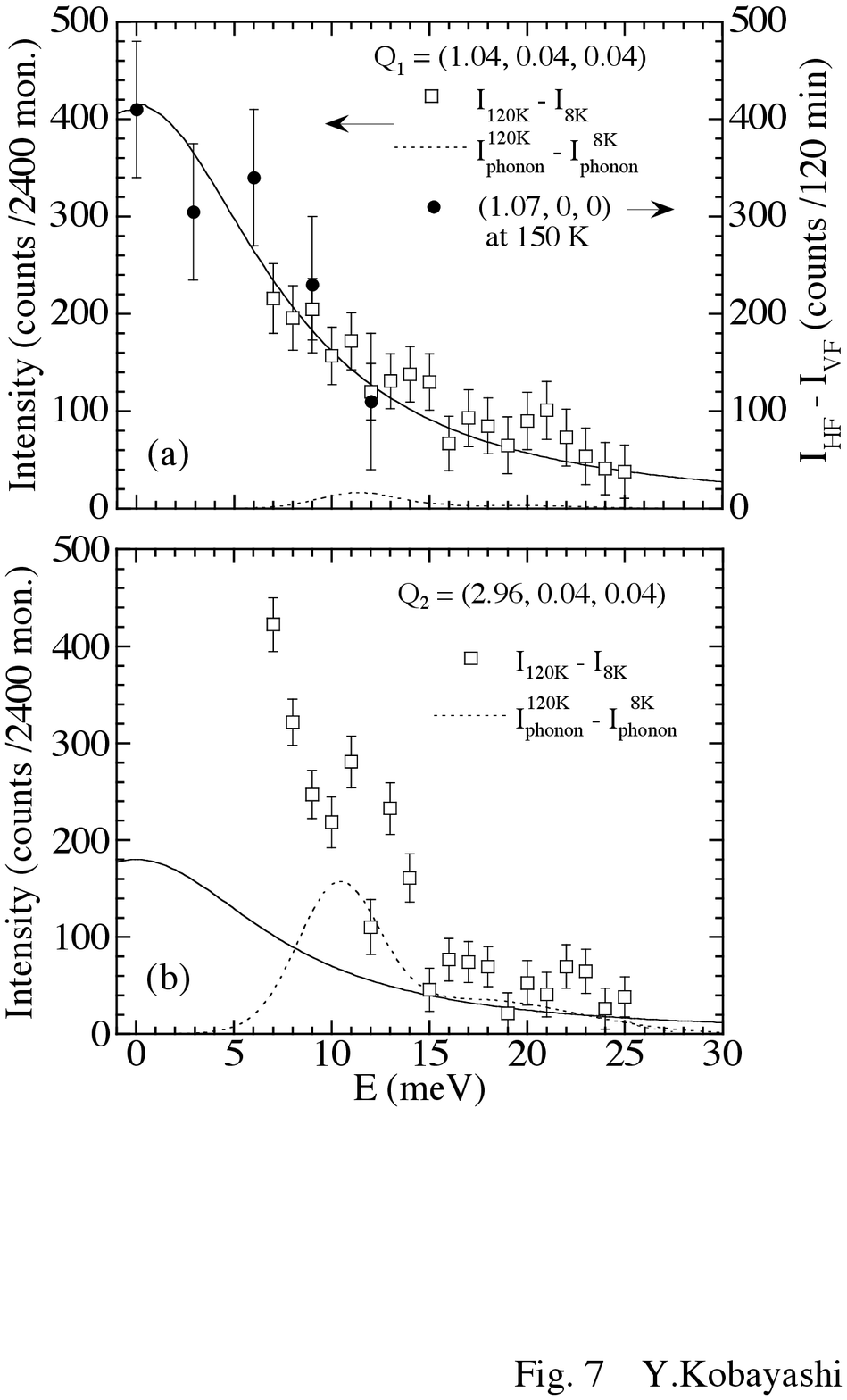}%
\caption{\label{fig:7}  The magnetic scattering intensity obtained by 
I(120 K) - I(8 K) measured at (a) $\bm{Q}_{1}$ and (b) $\bm{Q}_{2}$. 
Polarized neutron scattering data \cite{Asai1} are also shown.   The
solid lines denote the resolution-convoluted scattering function.
\cite{Asai1}  The dotted line shows the phonon scattering intensity.}  
\end{figure}

The difference of the neutron scattering between 8 and 120 K at 
$\bm{Q}_{1}$ = (1.04, 0.04, 0.04) is shown in Fig.~\ref{fig:7} (a) along
with the paramagnetic scattering intensity at (1.07, 0, 0) at 150 K
measured previously with polarized neutrons.\cite{Asai1}  Solid line
represents the magnetic scattering function  $\Bigl[ \ S(\bm{Q},\omega)
\propto \chi(\bm{Q}) [\omega \Gamma / (\omega^{2} +\Gamma^{2})] \{1/[1 -
\rm{exp}(- \omega/kT)]\} \ \Bigr]$ with $\Gamma = 4$ meV, convolved
with the instrumental resolution ($\Gamma_{res} = 5$ meV).  We see that
the increment of the scattering can be ascribed to the fluctuation of the
IS magnetic moments and is represented with the simple scattering function
up to energies as high as 25 meV, although the solid line is only a guide
to the eye.  Here the increment of the phonon intensity between 8 and 120
K (shown by dotted line) is negligible.  The difference of the neutron
scattering between 8 and 120 K at $\bm{Q}_{2}$ = (2.96, 0.04, 0.04)
[Fig.~\ref{fig:7} (b)] is also interpreted by the paramagnetic scattering
and increased intensity of the phonon scattering.  Here, the former and
the latter are denoted by the solid and dotted lines, respectively.  The
paramagnetic scattering intensity at $\bm{Q}_{2}$ is consistent with that
at $\bm{Q}_{1}$ after being normalized by the magnetic form factor.

\section{Conclusion}

We have investigated the phonons in LaCoO$_{3}$ using inelastic neutron 
scattering.  The phonon dispersion curves for 
high-symmetry directions and $E\lesssim25$~meV have been 
clarified.  For ($\delta$ $\delta$ $\delta$), both LA and TA phonon
branches show a symmetric shape about $\delta$ = 0.25.  For 
(0 $\delta$ $\delta$), the phonon energy of the TA branch levels off for
$\delta \gtrsim 0.3$.  These behaviors are consistent with the folded
Brillouin zone for the rhombohedral perovskite structure containing two
chemical formula units.  The transverse optical branches
continuing to the $E_{g}$ rotational mode of O atoms and the $E_{g}$
vibrational mode of La atoms show remarkable softening associated with the
spin-state transition.  The softening occurs over much of
the Brillouin zone.   In contrast, the acoustic phonons show a much
smaller change with temperature.  The magnetic scattering intensity at
around (1 0 0) increases to a maximum for $E \gtrsim 15$ meV, ascribable
to the excitation of Co$^{3+}$ from the LS to the IS spin-state.  At
higher temperatures, the paramagnetic scattering associated with the
population of the IS-state Co$^{3+}$ ions was observed up to 25 meV.

\begin{acknowledgments}
The authors thank A. Ishikawa and S. Sugai for fruitful discussion and for 
giving us their data before publication.  The authors also thank M.
Fujita and H. Hiraka for their help on the neutron scattering
experiments.  The authors also thank K. Abe for fruitful discussion. 
This work was supported by Grant-in-Aids for Scientific Research from the
Ministry of Education, Science, Sports and Culture of Japan, and from
UK-Japan Collaboration on Neutron Scattering.  This work was partially
supported by Foundation for Promotion of Material Science and Technology
of Japan.  Work at Brookhaven is supported by the Office of Science, U.S.
Department of Energy, under Contract No.\ DE-AC02-98CH10886.
\end{acknowledgments}


\begin{thebibliography}{25}
\expandafter\ifx\csname natexlab\endcsname\relax\def\natexlab#1{#1}\fi
\expandafter\ifx\csname bibnamefont\endcsname\relax
  \def\bibnamefont#1{#1}\fi
\expandafter\ifx\csname bibfnamefont\endcsname\relax
  \def\bibfnamefont#1{#1}\fi
\expandafter\ifx\csname citenamefont\endcsname\relax
  \def\citenamefont#1{#1}\fi
\expandafter\ifx\csname url\endcsname\relax
  \def\url#1{\texttt{#1}}\fi
\expandafter\ifx\csname urlprefix\endcsname\relax\def\urlprefix{URL }\fi
\providecommand{\bibinfo}[2]{#2}
\providecommand{\eprint}[2][]{\url{#2}}

\bibitem[{\citenamefont{Heikes et~al.}(1964)\citenamefont{Heikes, C.Miller, and
  Mazelsky}}]{Heikes}
\bibinfo{author}{\bibfnamefont{R.~R.} \bibnamefont{Heikes}},
  \bibinfo{author}{\bibfnamefont{R.}~\bibnamefont{C.Miller}}, \bibnamefont{and}
  \bibinfo{author}{\bibfnamefont{R.}~\bibnamefont{Mazelsky}},
  \bibinfo{journal}{Physica} \textbf{\bibinfo{volume}{30}},
  \bibinfo{pages}{1600} (\bibinfo{year}{1964}).

\bibitem[{\citenamefont{Raccah and Goodenough}(1967)}]{Raccah}
\bibinfo{author}{\bibfnamefont{P.~M.} \bibnamefont{Raccah}} \bibnamefont{and}
  \bibinfo{author}{\bibfnamefont{J.~B.} \bibnamefont{Goodenough}},
  \bibinfo{journal}{Phys. Rev.} \textbf{\bibinfo{volume}{155}},
  \bibinfo{pages}{932} (\bibinfo{year}{1967}).

\bibitem[{\citenamefont{Asai et~al.}(1989)\citenamefont{Asai, Gehring, Chou,
  and Shirane}}]{Asai1}
\bibinfo{author}{\bibfnamefont{K.}~\bibnamefont{Asai}},
  \bibinfo{author}{\bibfnamefont{P.}~\bibnamefont{Gehring}},
  \bibinfo{author}{\bibfnamefont{H.}~\bibnamefont{Chou}}, \bibnamefont{and}
  \bibinfo{author}{\bibfnamefont{G.}~\bibnamefont{Shirane}},
  \bibinfo{journal}{Phys. Rev. B} \textbf{\bibinfo{volume}{40}},
  \bibinfo{pages}{10982} (\bibinfo{year}{1989}).

\bibitem[{\citenamefont{Asai et~al.}(1994)\citenamefont{Asai, Yokokura,
  Nishimori, Chou, Tranquada, Shirane, Higuchi, Okajima, and Kohn}}]{Asai2}
\bibinfo{author}{\bibfnamefont{K.}~\bibnamefont{Asai}},
  \bibinfo{author}{\bibfnamefont{O.}~\bibnamefont{Yokokura}},
  \bibinfo{author}{\bibfnamefont{N.}~\bibnamefont{Nishimori}},
  \bibinfo{author}{\bibfnamefont{H.}~\bibnamefont{Chou}},
  \bibinfo{author}{\bibfnamefont{J.~M.} \bibnamefont{Tranquada}},
  \bibinfo{author}{\bibfnamefont{G.}~\bibnamefont{Shirane}},
  \bibinfo{author}{\bibfnamefont{S.}~\bibnamefont{Higuchi}},
  \bibinfo{author}{\bibfnamefont{Y.}~\bibnamefont{Okajima}}, \bibnamefont{and}
  \bibinfo{author}{\bibfnamefont{K.}~\bibnamefont{Kohn}},
  \bibinfo{journal}{Phys. Rev. B} \textbf{\bibinfo{volume}{50}},
  \bibinfo{pages}{3025} (\bibinfo{year}{1994}).

\bibitem[{\citenamefont{Asai et~al.}(1997)\citenamefont{Asai, Yokokura, Suzuki,
  Naka, Matsumoto, Takahashi, Mori, and Kohn}}]{Asai4}
\bibinfo{author}{\bibfnamefont{K.}~\bibnamefont{Asai}},
  \bibinfo{author}{\bibfnamefont{O.}~\bibnamefont{Yokokura}},
  \bibinfo{author}{\bibfnamefont{M.}~\bibnamefont{Suzuki}},
  \bibinfo{author}{\bibfnamefont{T.}~\bibnamefont{Naka}},
  \bibinfo{author}{\bibfnamefont{T.}~\bibnamefont{Matsumoto}},
  \bibinfo{author}{\bibfnamefont{H.}~\bibnamefont{Takahashi}},
  \bibinfo{author}{\bibfnamefont{N.}~\bibnamefont{Mori}}, \bibnamefont{and}
  \bibinfo{author}{\bibfnamefont{K.}~\bibnamefont{Kohn}}, \bibinfo{journal}{J.
  Phys. Soc. Jpn.} \textbf{\bibinfo{volume}{66}}, \bibinfo{pages}{967}
  (\bibinfo{year}{1997}).

\bibitem[{\citenamefont{Asai et~al.}(1998)\citenamefont{Asai, Yoneda, Yokokura,
  Tranquada, Shirane, and Kohn}}]{Asai3}
\bibinfo{author}{\bibfnamefont{K.}~\bibnamefont{Asai}},
  \bibinfo{author}{\bibfnamefont{A.}~\bibnamefont{Yoneda}},
  \bibinfo{author}{\bibfnamefont{O.}~\bibnamefont{Yokokura}},
  \bibinfo{author}{\bibfnamefont{J.~M.} \bibnamefont{Tranquada}},
  \bibinfo{author}{\bibfnamefont{G.}~\bibnamefont{Shirane}}, \bibnamefont{and}
  \bibinfo{author}{\bibfnamefont{K.}~\bibnamefont{Kohn}}, \bibinfo{journal}{J.
  Phys. Soc. Jpn.} \textbf{\bibinfo{volume}{67}}, \bibinfo{pages}{290}
  (\bibinfo{year}{1998}).

\bibitem[{\citenamefont{Baier et~al.}(2005)\citenamefont{Baier, Jodlauk,
  Kriener, Reichl, Zobel, Kierspel, Freimuth, and Lorenz}}]{baie05}
\bibinfo{author}{\bibfnamefont{J.}~\bibnamefont{Baier}},
  \bibinfo{author}{\bibfnamefont{S.}~\bibnamefont{Jodlauk}},
  \bibinfo{author}{\bibfnamefont{M.}~\bibnamefont{Kriener}},
  \bibinfo{author}{\bibfnamefont{A.}~\bibnamefont{Reichl}},
  \bibinfo{author}{\bibfnamefont{C.}~\bibnamefont{Zobel}},
  \bibinfo{author}{\bibfnamefont{H.}~\bibnamefont{Kierspel}},
  \bibinfo{author}{\bibfnamefont{A.}~\bibnamefont{Freimuth}}, \bibnamefont{and}
  \bibinfo{author}{\bibfnamefont{T.}~\bibnamefont{Lorenz}},
  \bibinfo{journal}{Phys. Rev. B.} \textbf{\bibinfo{volume}{71}},
  \bibinfo{pages}{014443} (\bibinfo{year}{2005}).

\bibitem[{\citenamefont{Nekrasov et~al.}(2003)\citenamefont{Nekrasov,
  Streltsov, Korotin, and Anisimov}}]{nekr03}
\bibinfo{author}{\bibfnamefont{I.~A.} \bibnamefont{Nekrasov}},
  \bibinfo{author}{\bibfnamefont{S.~V.} \bibnamefont{Streltsov}},
  \bibinfo{author}{\bibfnamefont{M.~A.} \bibnamefont{Korotin}},
  \bibnamefont{and} \bibinfo{author}{\bibfnamefont{V.~I.}
  \bibnamefont{Anisimov}}, \bibinfo{journal}{Phys. Rev. B.}
  \textbf{\bibinfo{volume}{68}}, \bibinfo{pages}{235113}
  (\bibinfo{year}{2003}).

\bibitem[{\citenamefont{Korotin et~al.}(1996)\citenamefont{Korotin, Ezhov,
  Solovyev, Anisimov, Khomskii, and Sawatzky}}]{Korotin}
\bibinfo{author}{\bibfnamefont{M.~A.} \bibnamefont{Korotin}},
  \bibinfo{author}{\bibfnamefont{S.~Y.} \bibnamefont{Ezhov}},
  \bibinfo{author}{\bibfnamefont{I.~V.} \bibnamefont{Solovyev}},
  \bibinfo{author}{\bibfnamefont{V.~I.} \bibnamefont{Anisimov}},
  \bibinfo{author}{\bibfnamefont{D.~I.} \bibnamefont{Khomskii}},
  \bibnamefont{and} \bibinfo{author}{\bibfnamefont{G.~A.}
  \bibnamefont{Sawatzky}}, \bibinfo{journal}{Phys. Rev. B}
  \textbf{\bibinfo{volume}{54}}, \bibinfo{pages}{5309} (\bibinfo{year}{1996}).

\bibitem[{\citenamefont{Tanabe and Sugano}(1954)}]{Tanabe}
\bibinfo{author}{\bibfnamefont{Y.}~\bibnamefont{Tanabe}} \bibnamefont{and}
  \bibinfo{author}{\bibfnamefont{S.}~\bibnamefont{Sugano}},
  \bibinfo{journal}{J. Phys. Soc. Jpn.} \textbf{\bibinfo{volume}{9}},
  \bibinfo{pages}{766} (\bibinfo{year}{1954}).

\bibitem[{\citenamefont{Radwa\'nski and Ropka}(1999)}]{radw99}
\bibinfo{author}{\bibfnamefont{R.~J.} \bibnamefont{Radwa\'nski}}
  \bibnamefont{and} \bibinfo{author}{\bibfnamefont{Z.}~\bibnamefont{Ropka}},
  \bibinfo{journal}{Solid State Commun.} \textbf{\bibinfo{volume}{112}},
  \bibinfo{pages}{621} (\bibinfo{year}{1999}).

\bibitem[{\citenamefont{Noguchi et~al.}(2002)\citenamefont{Noguchi, Kawamata,
  Okuda, Nojiri, and Motokawa}}]{Noguchi}
\bibinfo{author}{\bibfnamefont{S.}~\bibnamefont{Noguchi}},
  \bibinfo{author}{\bibfnamefont{S.}~\bibnamefont{Kawamata}},
  \bibinfo{author}{\bibfnamefont{K.}~\bibnamefont{Okuda}},
  \bibinfo{author}{\bibfnamefont{H.}~\bibnamefont{Nojiri}}, \bibnamefont{and}
  \bibinfo{author}{\bibfnamefont{M.}~\bibnamefont{Motokawa}},
  \bibinfo{journal}{Phys. Rev. B} \textbf{\bibinfo{volume}{66}},
  \bibinfo{pages}{94404} (\bibinfo{year}{2002}).

\bibitem[{\citenamefont{Murata et~al.}(1999)\citenamefont{Murata, Ishida,
  Suzuki, Kobayashi, Asai, and Kohn}}]{Murata}
\bibinfo{author}{\bibfnamefont{S.}~\bibnamefont{Murata}},
  \bibinfo{author}{\bibfnamefont{S.}~\bibnamefont{Ishida}},
  \bibinfo{author}{\bibfnamefont{M.}~\bibnamefont{Suzuki}},
  \bibinfo{author}{\bibfnamefont{Y.}~\bibnamefont{Kobayashi}},
  \bibinfo{author}{\bibfnamefont{K.}~\bibnamefont{Asai}}, \bibnamefont{and}
  \bibinfo{author}{\bibfnamefont{K.}~\bibnamefont{Kohn}},
  \bibinfo{journal}{Physica B} \textbf{\bibinfo{volume}{263-264}},
  \bibinfo{pages}{647} (\bibinfo{year}{1999}).

\bibitem[{\citenamefont{Yamaguchi et~al.}(1997)\citenamefont{Yamaguchi,
  Okimoto, and Tokura}}]{Yamaguchi}
\bibinfo{author}{\bibfnamefont{S.}~\bibnamefont{Yamaguchi}},
  \bibinfo{author}{\bibfnamefont{Y.}~\bibnamefont{Okimoto}}, \bibnamefont{and}
  \bibinfo{author}{\bibfnamefont{Y.}~\bibnamefont{Tokura}},
  \bibinfo{journal}{Phys. Rev. B} \textbf{\bibinfo{volume}{55}},
  \bibinfo{pages}{R8666} (\bibinfo{year}{1997}).

\bibitem[{\citenamefont{Louca and Sarrao}(2003)}]{Louca}
\bibinfo{author}{\bibfnamefont{D.}~\bibnamefont{Louca}} \bibnamefont{and}
  \bibinfo{author}{\bibfnamefont{J.~L.} \bibnamefont{Sarrao}},
  \bibinfo{journal}{Phys. Rev. Lett} \textbf{\bibinfo{volume}{91}},
  \bibinfo{pages}{155501} (\bibinfo{year}{2003}).

\bibitem[{\citenamefont{Maris et~al.}(2003)\citenamefont{Maris, Ren,
  Volotchaev, Zobel, Lorenz, and Palstra}}]{Maris}
\bibinfo{author}{\bibfnamefont{G.}~\bibnamefont{Maris}},
  \bibinfo{author}{\bibfnamefont{Y.}~\bibnamefont{Ren}},
  \bibinfo{author}{\bibfnamefont{V.}~\bibnamefont{Volotchaev}},
  \bibinfo{author}{\bibfnamefont{C.}~\bibnamefont{Zobel}},
  \bibinfo{author}{\bibfnamefont{T.}~\bibnamefont{Lorenz}}, \bibnamefont{and}
  \bibinfo{author}{\bibfnamefont{T.~T.~M.} \bibnamefont{Palstra}},
  \bibinfo{journal}{Phys. Rev. B} \textbf{\bibinfo{volume}{67}},
  \bibinfo{pages}{224423} (\bibinfo{year}{2003}).

\bibitem[{\citenamefont{Kanamori}(1960)}]{Kanamori}
\bibinfo{author}{\bibfnamefont{J.}~\bibnamefont{Kanamori}},
  \bibinfo{journal}{J. Appl. Phys.} \textbf{\bibinfo{volume}{31}},
  \bibinfo{pages}{14S} (\bibinfo{year}{1960}).

\bibitem[{\citenamefont{Ishikawa et~al.}(2004)\citenamefont{Ishikawa, Nomura,
  and Sugai}}]{Ishikawa}
\bibinfo{author}{\bibfnamefont{A.}~\bibnamefont{Ishikawa}},
  \bibinfo{author}{\bibfnamefont{J.}~\bibnamefont{Nohara}}, \bibnamefont{and}
  \bibinfo{author}{\bibfnamefont{S.}~\bibnamefont{Sugai}},
  \bibinfo{journal}{Phys. Rev. Lett} \textbf{\bibinfo{volume}{93}},
  \bibinfo{pages}{136401} (\bibinfo{year}{2004}).

\bibitem[{\citenamefont{Axe et~al.}(1969)\citenamefont{Axe, Shirane, and
  Muller}}]{Axe}
\bibinfo{author}{\bibfnamefont{J.~D.} \bibnamefont{Axe}},
  \bibinfo{author}{\bibfnamefont{G.}~\bibnamefont{Shirane}}, \bibnamefont{and}
  \bibinfo{author}{\bibfnamefont{K.~A.} \bibnamefont{Muller}},
  \bibinfo{journal}{Phys. Rev.} \textbf{\bibinfo{volume}{183}},
  \bibinfo{pages}{820} (\bibinfo{year}{1969}).

\bibitem[{\citenamefont{Kiems et~al.}(1973)\citenamefont{Kiems, Shirane,
  Muller, and Scheel}}]{Kiems}
\bibinfo{author}{\bibfnamefont{J.~K.} \bibnamefont{Kiems}},
  \bibinfo{author}{\bibfnamefont{G.}~\bibnamefont{Shirane}},
  \bibinfo{author}{\bibfnamefont{K.~A.} \bibnamefont{Muller}},
  \bibnamefont{and} \bibinfo{author}{\bibfnamefont{H.~J.}
  \bibnamefont{Scheel}}, \bibinfo{journal}{Phys. Rev. B}
  \textbf{\bibinfo{volume}{8}}, \bibinfo{pages}{1119} (\bibinfo{year}{1973}).

\bibitem[{\citenamefont{Kobayashi et~al.}(2000)\citenamefont{Kobayashi,
  Fujiwara, Murata, Asai, and Yasuoka}}]{Kobayashi1}
\bibinfo{author}{\bibfnamefont{Y.}~\bibnamefont{Kobayashi}},
  \bibinfo{author}{\bibfnamefont{N.}~\bibnamefont{Fujiwara}},
  \bibinfo{author}{\bibfnamefont{S.}~\bibnamefont{Murata}},
  \bibinfo{author}{\bibfnamefont{K.}~\bibnamefont{Asai}}, \bibnamefont{and}
  \bibinfo{author}{\bibfnamefont{H.}~\bibnamefont{Yasuoka}},
  \bibinfo{journal}{Phys. Rev. B} \textbf{\bibinfo{volume}{62}},
  \bibinfo{pages}{410} (\bibinfo{year}{2000}).

\bibitem[{\citenamefont{Thornton et~al.}(1986)\citenamefont{Thornton, Tofield,
  and Hewat}}]{Thornton2}
\bibinfo{author}{\bibfnamefont{G.}~\bibnamefont{Thornton}},
  \bibinfo{author}{\bibfnamefont{B.~C.} \bibnamefont{Tofield}},
  \bibnamefont{and} \bibinfo{author}{\bibfnamefont{A.~W.} \bibnamefont{Hewat}},
  \bibinfo{journal}{J. Solid State Chem.} \textbf{\bibinfo{volume}{61}},
  \bibinfo{pages}{301} (\bibinfo{year}{1986}).

\bibitem[{\citenamefont{Granado et~al.}(1998)\citenamefont{Granado, Moreno,
  Garcia, Sanjurjo, Rettori, Torriani, Oseroff, Neumeier, McClellan, Cheong
  et~al.}}]{Granado}
\bibinfo{author}{\bibfnamefont{E.}~\bibnamefont{Granado}},
  \bibinfo{author}{\bibfnamefont{N.~O.} \bibnamefont{Moreno}},
  \bibinfo{author}{\bibfnamefont{A.}~\bibnamefont{Garcia}},
  \bibinfo{author}{\bibfnamefont{J.~A.} \bibnamefont{Sanjurjo}},
  \bibinfo{author}{\bibfnamefont{C.}~\bibnamefont{Rettori}},
  \bibinfo{author}{\bibfnamefont{I.}~\bibnamefont{Torriani}},
  \bibinfo{author}{\bibfnamefont{S.~B.} \bibnamefont{Oseroff}},
  \bibinfo{author}{\bibfnamefont{J.~J.} \bibnamefont{Neumeier}},
  \bibinfo{author}{\bibfnamefont{K.~J.} \bibnamefont{McClellan}},
  \bibinfo{author}{\bibfnamefont{S.~W.} \bibnamefont{Cheong}},
  \bibnamefont{et~al.}, \bibinfo{journal}{Phys. Rev. B}
  \textbf{\bibinfo{volume}{58}}, \bibinfo{pages}{11435} (\bibinfo{year}{1998}).

\bibitem[{\citenamefont{Abrashev et~al.}(1999)\citenamefont{Abrashev,
  Litvinchuk, Iliev, Meng, Popov, Ivanov, Chakalov, and Thomsen}}]{Abrashev}
\bibinfo{author}{\bibfnamefont{M.~V.} \bibnamefont{Abrashev}},
  \bibinfo{author}{\bibfnamefont{A.~P.} \bibnamefont{Litvinchuk}},
  \bibinfo{author}{\bibfnamefont{M.~N.} \bibnamefont{Iliev}},
  \bibinfo{author}{\bibfnamefont{R.~L.} \bibnamefont{Meng}},
  \bibinfo{author}{\bibfnamefont{V.~N.} \bibnamefont{Popov}},
  \bibinfo{author}{\bibfnamefont{V.~G.} \bibnamefont{Ivanov}},
  \bibinfo{author}{\bibfnamefont{R.~A.} \bibnamefont{Chakalov}},
  \bibnamefont{and} \bibinfo{author}{\bibfnamefont{C.}~\bibnamefont{Thomsen}},
  \bibinfo{journal}{Phys. Rev. B} \textbf{\bibinfo{volume}{59}},
  \bibinfo{pages}{4146} (\bibinfo{year}{1999}).

\bibitem[{\citenamefont{Stirling}(1972)}]{Stirling}
\bibinfo{author}{\bibfnamefont{W.~G.} \bibnamefont{Stirling}},
  \bibinfo{journal}{J. Phys. C: Solid State Phys.}
  \textbf{\bibinfo{volume}{5}}, \bibinfo{pages}{2711} (\bibinfo{year}{1972}).

\end{thebibliography}

\end{document}